# Hosting exceptional points in 1D photonic bandgap waveguide for mode engineering

**Sibnath Dey and Somnath Ghosh\***

*Unconventional Photonics Laboratory, Department of Physics, Indian Institute of Technology Jodhpur, Rajasthan-342037, India*

*Emails: somiit@rediffmail.com\**

**Abstract:** We report a planar 1D photonic bandgap waveguide exhibiting four second-order exceptional points (EP2s). The interactions between the selective pairs of supported quasi-guided TE modes are modulated by spatial distribution of transverse in-homogeneous gain-loss profile.

## 1. Introduction

Non-Hermitian quantum mechanics forms the basis in realizing open photonic systems where non-Hermiticity is achieved by optical gain-loss. One of the exclusive non-Hermitian features is appearance of exceptional points (EPs) where at least two coupled eigenvalues and the corresponding eigenstates simultaneously coalesce [1, 2]. In the photonic system, the appearance of EPs can be exploited to a broad range of interesting applications, including lasing [3], asymmetric mode switching [1, 4], nonreciprocal light transmission [5], and ultrasensitive sensing [6]. A second-order EP (EP2) can be considered a topological branch point singularity, leading to significant modification in the system's behavior. The unconventional behavior of the two eigenmodes can be observed with the continuous exchange between them following an adiabatic parametric encirclement while enclosing an EP2. On the other hand, the photonic bandgap waveguide offers a full range of opportunities to control light in the system. The need still exists for low-loss waveguides that offer control over guiding properties such as effective index tunability, mode size, group velocity dispersions, etc. Bragg Reflection Waveguide (BRW) [7] has emerged as a strong contender for various optical applications due to the extra degrees-of-freedom for choosing parameters to control light. It is also expected that utilizing Bragg waveguiding in the transverse direction can introduce additional freedom in modifying the dispersive properties as well as modal behavior of the waveguide. In this paper, we propose a new prototype of gain-loss assisted Bragg refection waveguide (BRW) that supports nine bandgap guided transverse electric (TE) modes. We first time introduce the non-Hermiticity in the BRW structure with spatial variation of transverse inhomogeneous gain-loss profile. Here, we control the interactions between the specific bandgap guided modes by proper tuning of two independent parameters viz. gain coefficient ($\gamma$) and loss-to-gain ratio ($\tau$). We have identified four EP2s among the eight bandgap guided modes in the system parameter space by proper tuning of active parameters ($\gamma$ and $\tau$). Interestingly, we observed that the interaction between any pair of modes is not affected by the others, which essentially reflects that four identified EP2s (i.e., $EP2^{(1)}$, $EP2^{(2)}$, $EP2^{(3)}$, $EP2^{(4)}$) are independent and noninteracting. We have also presented four unique second-order flip-of-states phenomena by considering a quasi-static parametric encirclement around the four identified EP2s individually in the ($\gamma, \tau$)-plane. We believe that the existence of the EP2s in the system parameter space and their promising applications in the photonic bandgap guided structure may open up new possibilities in the next-generation photonic integrated circuits.

## 2. Results and discussions

We have designed a BRW structure consist of a low index core (r.i.: $n_{co} = 1.458$, width: $d_{co} = 6$ μm), sandwiched between identical finite periodic layers, which is shown in Fig.1.(a). Our 1D bandgap guided structure consists of three periodic layers in both side, where the spatial widths of high index ($n_2 = 3.927$) and low index ($n_1 = 2.15$) layers have been chosen to be 0.1 μm ($= d_2$) and 0.135 μm ($= d_1$) respectively. A schematic of the complex refractive index profile for specific $\gamma = 0.035$ and $\tau = 1.5$ have been shown in the Fig.1.(b). We have chosen three periodic layers on both sides of the core to confirm maximum confinement of the eight bandgap guided modes in the core region. The chosen number of the bilayers results in low leakage loss in the system. To guide the light of wavelength ($\lambda$)1.55 $\mu m$ using the BRW, bandgap engineering has been performed to realize the Bragg stopband by modulating the spatial periodicity ($\Lambda = d_1 + d_2$) and corresponding refractive indices ($n_1$ and $n_2$) of chosen periodic layers. We have chosen higher refractive index difference that confirms a wider photonic bandgap, as shown in Fig.1.(c). The parameters are chosen in such a way that waveguide supports nine quasi-guided TE modes (i.e., $TE_i$; {$i = 0 ... 8$}). But we have shown eight bandgap guided modes in the Fig.1.(d). In the Fig.1.(d.1), we have shown the field confinements of $TE_0$ (red curve) and $TE_1$ (black curve). Similarly in Fig.1.(d.2), we have shown the field confinements of $TE_2$ (blue curve) and $TE_3$ (green curve). Similarly in Fig.1.(d.3) and Fig.1.(d.4) indicate the confinement of the pairs of ($TE_4$ and $TE_5$) and ($TE_6$ and $TE_7$), respectively. Here we ignore the presence of transverse magnetic (TM) modes due to relative higher leakage loss. The interaction among the bandgap guided modes with one to one coupling restriction is controlled by the spatial distribution of transverse inhomogeneous gain-loss profile. Considering the optimized operating parameters with proper tuning of unbalanced gain-loss profile in the core region, we have identified four EP2s (Fig.2) in $(\gamma, \tau)$-plane using the concept of Avoided resonance crossing (ARC) [1, 2]. We confirm the presence of four different EP2s in $(\gamma, \tau)$-plane when complex $n_{eff}$-values associated with coupled modes coalesce in the complex $n_{eff}$-plane. We investigate the ARC phenomena between two specified modes with crossing/anticrossing of their real and imaginary parts (i.e., Re[$n_{eff}$] and Im[$n_{eff}$] for two different $\tau$-values) respectively, in specified $\gamma$-range. We investigate several cases and observed that for $\tau = 1.5$, $n_{eff0}$ and $n_{eff1}$ coalesce near $\gamma = 0.0086$ as can be seen in Fig.2.(a.1). Thus, we have numerically detected the position of an EP2 at $(\gamma_{EP} = 0.0086, \tau_{EP} = 1.5)$ which has been denoted as EP2$^{(1)}$. Similarly, we observed that $n_{eff2}$ and $n_{eff3}$ coalesce near $\gamma = 0.0248$ as can be seen in Fig.2.(a.2). Thus, we numerically detect the position of other EP2 at $(\gamma_{EP} = 0.0248, \tau_{EP} = 1.46)$, which has been denoted as EP2$^{(2)}$.

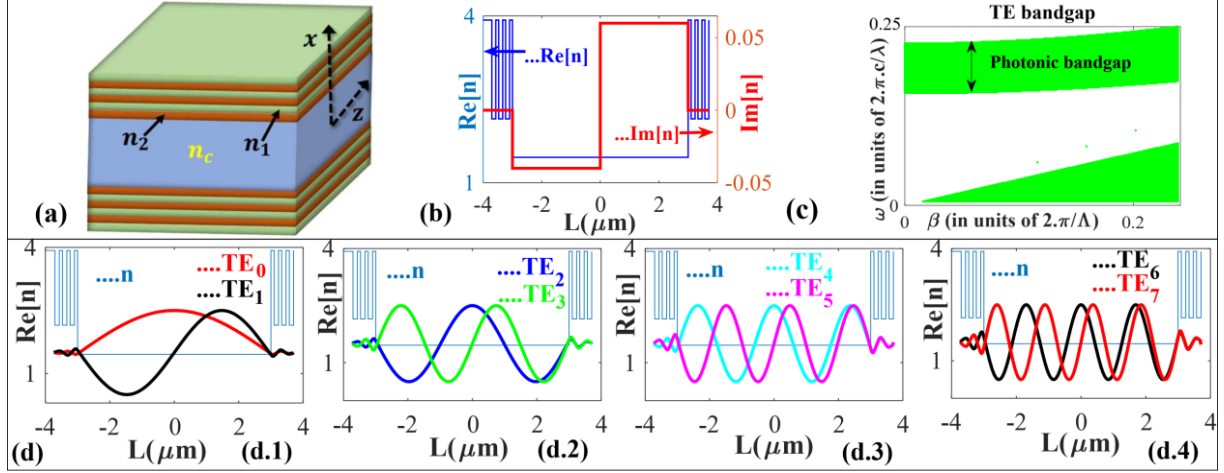

**Fig. 1:** (a) Schematic of a planar Bragg reflection waveguide ($x \to transverse\ axis$, $z \to propagation\ axis$). (b) Transverse complex refractive index profile, where Re[n] (blue line) and Im[n] (red line; for $\gamma = 0.035$, and $\tau = 1.5$) represents the real and imaginary parts of the refractive index profiles, respectively. (c) Optimized band diagram (TE polarization mode) for proposed periodic design. (d) Optimized real part of the refractive index (blue line) of the waveguide and quasi guided TE$_i$ (i = 0,1,2,3,4,5,6,7) polarization modes. (d.1) represents for TE$_0$ (red curve) and TE$_1$ (black curve), (d.2) represents for TE$_2$ (blue curve) and TE$_3$ (green curve), (d.3) represents for TE$_4$ (cyan curve) and TE$_5$ (pink curve) and (d.4) represents for TE$_6$ (black curve) and TE$_7$ (red curve).

Similarly, the position of another two EP2s at $(\gamma_{EP} = 0.0459, \tau_{EP} = 1.34)$ and $(\gamma_{EP} = 0.0698, \tau_{EP} = 1.25)$ are identified, which has been denoted as EP2$^{(3)}$ [Fig.2.(a.3)] and EP2$^{(4)}$ [Fig.2.(a.4)], respectively.

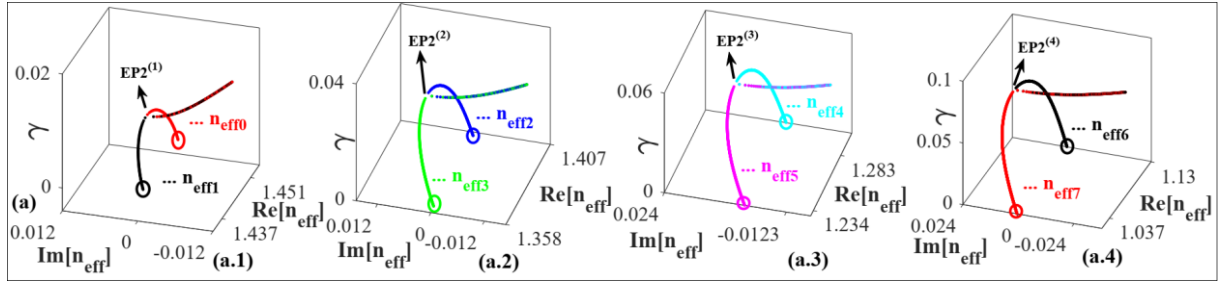

**Fig. 2:** (a) Encounter of EP2s; (a.1) Dynamics of complex $n_{eff0}$ (dotted red line) and $n_{eff1}$ (dotted black line) with respect to $\gamma$ for chosen $\tau = 1.5$, where the modes coalescence at EP2$^{(1)}$ near $\gamma = 0.0086$. Similarly, (a.2) indicates the dynamics of complex $n_{eff2}$ (dotted blue line) and $n_{eff3}$ (dotted green line) for a chosen $\tau = 1.46$, showing coalescence at EP2$^{(2)}$ near $\gamma = 0.0248$. (a.3) indicates the dynamics of complex $n_{eff4}$ (dotted cyan line) and $n_{eff5}$ (dotted pink line) for chosen $\tau = 1.34$, where modes coalescence at EP2$^{(3)}$ near $\gamma = 0.0459$ (a.4) Indicates the dynamics of complex $n_{eff6}$ (dotted black line) and $n_{eff7}$ (dotted red line) for a chosen $\tau = 1.25$ showing coalescence at EP2$^{(4)}$ near $\gamma = 0.0698$.

The second-order branch point behavior of all the identified EP2s has been verified by encircling it in $(\gamma, \tau)$-plane. In the Fig.3.(a), to investigate the EP2s- encirclement scheme, we consider a closed elliptical parametric loop in $(\gamma, \tau)$-plane following the coupled equation $\gamma(\phi) = \gamma_0 \sin(\phi/2)$ and $\tau(\phi) = \tau_{EP} + r \sin\phi$ ($r$ is a characteristics parameter, and $\phi$ is tunable angle within $[0, 2\pi]$ that governs the closed variation of $\gamma$ and $\tau$). For $\gamma_0 < \gamma_{EP}$, the parameter loop does not enclose the specific EP2. Such type of encirclement process allows to reach $\gamma = 0$ at both the beginning and the end. So, it is necessary for device application point of view where one should find clean passive modes at the end of the encirclement process. Now we judiciously set the encirclement parameters ($\gamma_0 = 0.01$, $\tau_{EP} = 1.5$, and $r = 0.05$) to enclose only EP2$^{(1)}$ which has been shown in the black curve in Fig.3.(a). Here, we have observed that for one round parametric encirclement the only coupled pair ($n_{eff0}$ and $n_{eff1}$) of the corresponding TE (TE$_0$ and TE$_1$) modes mutually exchange their initial positions (flip-of-states) which is shown in Fig.3.(b.1). Now we set the another encirclement parameters to enclose EP2$^{(2)}$ which is shown in the red curve in Fig.3.(a). In this case, we have observed that the coupled pair ($n_{eff2}$ and $n_{eff3}$) of the corresponding TE

(TE$_2$ and TE$_3$) modes mutually exchange their initial positions, as shown in Fig.3.(b.2). Similarly in Fig.3.(b.3), we have observed another flip-of-state phenomenon between the coupled pair ($n_{\text{eff}4}$ and $n_{\text{eff}5}$) of the corresponding TE (TE$_4$ and TE$_5$) modes. Similarly, if we have encircle EP2$^{(4)}$ by setting the specific encirclement parameters, we have also observed that the coupled pair ($n_{\text{eff}6}$ and $n_{\text{eff}7}$) of the corresponding TE (TE$_6$ and TE$_7$) modes mutually exchange their initial positions, as shown in Fig.3.(b.4). We have also observed that the flip-of-states phenomena is topological because it is independent of the shape of the parameters space. Irrespective of the shape of the encirclement loop, we have observed the swapping of modes in the complex $n_{\text{eff}}$-plane. Here four identified EP2s are non-interacting and form an exceptional line in the parameter space. Thus we observed that the mode flipping phenomena owing to the specific EP2 ignoring the annoying influence of nearby modes.

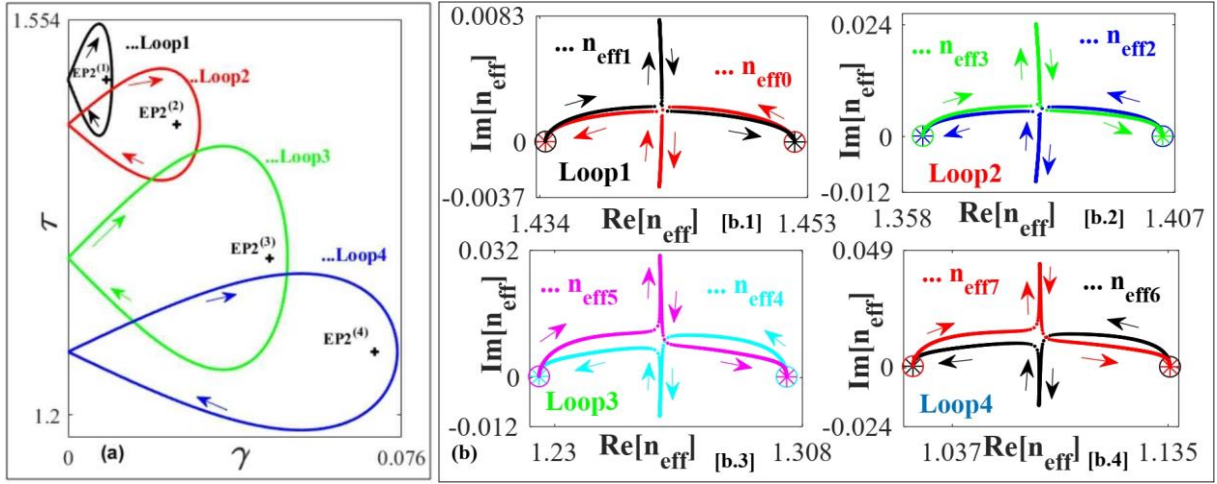

**Fig. 3:** (a) Chosen parameter loops in ($\gamma, \tau$)-plane to encircle individual EP2s, where black loop indicates the encirclement of EP2$^{(1)}$, red loop indicates the encirclement of EP2$^{(2)}$, green loop indicates the encirclement of EP2$^{(3)}$ and blue loop indicates the encirclement of EP2$^{(4)}$. (b) Trajectories of complex $n_{\text{eff}i}$ (i = 0,1,3,4,5,6,7) (represented by dotted red, black, blue, green, cyan, pink, black, red) in complex $n_{\text{eff}i}$-plane following the adiabatic parametric variation in ($\gamma, \tau$)-plane along (b.1) Loop1, indicates the adiabatic switching between $n_{\text{eff}0}$ and $n_{\text{eff}1}$, un-affecting others; (b.2) Loop2, indicates the adiabatic mode exchange between $n_{\text{eff}2}$ and $n_{\text{eff}3}$, un-affecting others; (b.3) Loop3, indicates the adiabatic mode exchange between $n_{\text{eff}4}$ and $n_{\text{eff}5}$, un-affecting others; and (b.4) Loop4, indicates the adiabatic mode exchange between $n_{\text{eff}6}$ and $n_{\text{eff}7}$, un-affecting others. Un-affected trajectories related to respective loops are not shown in the figures. Circular and star marker of the respective colors represent their initial and final position respectively (i.e., for $\phi = 0$). Arrows of respective colors specify their direction of progressions.

## 3. Summary

In summary, here we report for the first time a 1D Bragg refection waveguide with transverse distribution of inhomogeneous gain-loss profile. The waveguide hosts four non-interacting EP2s. The consecutive bandgap guided TE modes are coupled with one to one coupling restrictions and are analytically connected by four identified EP2s. The waveguide hosts four specific parameter spaces to encircle four EP2s individually. As long as the EP2s are encircled in the parameter space, flip-of-states phenomena is omnipresent. The second-order branch point behavior of the four identified EP2s has been revealed in terms of successive mode-switching between two specific modes following a quasi-static parametric encirclement process around each EP2s. The proposed photonic bandgap structure provides extra degrees of freedom to choose parameters that help in exploring the various properties of EP2s and their impact on transmission and reflection spectra. Our proposed scheme can open up new possibilities to boost the device applications associated with next-generation all-optical communication and computing.

**Funding:** S.D. acknowledges the financial support from the Ministry of Education, Government of India. S.G. acknowledges the financial support from the Science and Engineering Research Board (SERB) (Grant No. ECR/2017/000491), Government of India.

## 4. References

[1] S. Ghosh, and Y. D. Chong, "Exceptional points and asymmetric mode conversion in quasi-guided dual-mode optical waveguides," Sci. Rep **6**, 19837 (2016).
[2] S. Dey, A. Laha, and S. Ghosh, "Nonadiabatic modal dynamics around a third order Exceptional Point in a planar waveguide," Opt. Comm. **483**, 126644 (2020).
[3] L. Feng, Z. J. Wong, R.-M. Ma, Y. Wang, and X. Zhang, "Single mode laser by parity-time symmetry breaking," Science **346**, 972(2014)
[4] S. Dey, A. Laha, and S. Ghosh, "Nonlinearity-induced anomalous mode collapse and nonchiral asymmetric mode switching around multiple exceptional points," Phys. Rev. B **101**, 125432 (2020).
[5] A. Laha, S. Dey, H. K Gandhi, A. Biswas, and S. Ghosh, "Exceptional point and toward mode selective optical isolation," ACS Photonics **7**, 967 (2020).
[6] J. Wiersig, "Sensors operating at exceptional points: General theory," Phys. Rev. A **93**, 033809 (2016).
[7] B. R. West and A. S. Helmy, "Properties of the quarter-wave Bragg reflection waveguide: theory," J. Opt. Soc. Am. B **23,** 1207 (2006).